\documentstyle[12pt,aps,epsf]{revtex}
\begin{document}
   \title{Relativistic Charge Form Factor of the Deuteron\\ 
   from (np)--Scattering Phase Shifts}
     \author{Andrei V. Afanasev\thanks{On leave from Kharkov Institute of Physics
and Technology, Kharkov, Ukraine}  \\ {\it North Carolina Central University,
   Durham, NC 27707, USA} \\ {\it and} \\ {\it Jefferson Lab,
   Newport News, VA 23606, USA}\\
 V.D. Afanasev, S.V. Trubnikov \\
{\it Kharkov State University, 310077 Kharkov, Ukraine} }
\date{January 20, 2000}
\maketitle

\begin{abstract}
      Relativistic integral representation in terms of experimental
neutron--proton scattering phase shifts   is used to compute the
charge form factor of the deuteron $ G_{Cd}(Q^2)$. The results of numerical
 calculations of $|G_{Cd}(Q^2)|$ are presented in the interval of the
four--momentum transfers squared \mbox{$ 0 \leq Q^2 \leq 35\:fm^{-2}$.}
Zero and the prominent secondary maximum in  $ |G_{Cd}(Q^2)|$ are the
direct consequences of the change of sign in the experimental $^3S_1
$-- phase shifts. Till the point $Q^2 \simeq 20 \;fm^{-2}$  the calculated total
relativistic correction to $ |G_{Cd}(Q^2)|$ is positive and reaches the
maximal value of 25\% at  $ Q^2 \simeq 14 fm^{-2}$.
\end{abstract}

\begin{sloppypar}
      Deuteron is the  brightest example of intersection of nuclear and
particle physics. During more than sixty years it serves as a source of
important information about the nuclear forces, mesonic and  baryonic
degrees of freedoms in nuclei, relativistic effects and a possible role of
 quarks in nuclear structure. Therefore it is not surprising that currently
the electromagnetic (EM) structure of the deuteron is a subject of
 intensive theoretical (the list of publication is immense) and
experimental investigations.

With new experimental data from Jefferson Lab on elastic electron-deuteron
scattering  \cite{betsy},\cite{makis} at
  momentum transfers in the GeV-range, one needs to develop relativistic
 approaches to the (np)-bound state problem.  Recent experimental results
 from MIT-Bates \cite{bates} provided the first experimental evidence for a
 zero in the deuteron charge form factor $G_{Cd}$  at about $Q^2$= 20
fm$^{-2}$ predicted in a number of theoretical models (or not predicted, as
in some kinds of quark models).

      Here we report the new results of numerical calculations of
relativistic deuteron charge form factor (RDCFF). These calculations are
based on the approach to the relativistic impulse  approximation, which was
briefly discussed in ref.~\cite{tro} (see also the review~\cite{muza}
and, especially, the references therein).  The more detailed formulae are
contained in ref.~\cite{roman}.  In this approach the deuteron electromagnetic form factors 
(EMFF) are expressed
in terms of experimental neutron--proton $(n-p)$ phase shifts in the
triplet channel of $n-p$ scattering and nucleon
EMFFs. The approach was developed during a number of years and is still being
developed. The key components are as follows.

\noindent 1. General relativistic parametrization of the matrix elements of
the EM current between the n-p scattering states $\langle n'p'|j_\mu
|np\rangle $ in terms of the invariant EMFFs of the n-p system. This
parametrization is carried out in the canonical basis $|P\mu JlSm_J\rangle
$ of the two-particle state vectors. Here $P_\mu =(p_\mu+n_\mu )$ is the
total 4-momentum of n-p system; $J,m_J$ are the total angular moment and
its projection; $l,S$ are the orbital moment and total spin of n-p system.
As the result for the n-p scattering state in the elastic channel with
deuteron's quantum numbers ($J=S=J'=S'=1; l,l'=0,2$) we obtain sixteen
invariant FFs $G_i^{ll'}(s,s',t,...)\mbox{, where}\quad s \equiv (p_\mu
+n_\mu )^2,\; t=q^2_\mu  \equiv -Q^2 <0 \footnote{ For the  choice of
kinematical variables   see App.A.}$ and the index $i$ numerates the
charge(C), magnetic(M), quadrupole(Q) and the quadrupole of the second
generation, or toroidal(T) FFs.

\noindent 2. The decomposition of the total $G^{..}_{.} $ into the sum
\begin{equation}
G=g +G^{int},
\label{decomp}
\end{equation}
where $g^{..}_{..}$ is the FFs of unconnected part of the current (see
Fig.1).  This part is defined as usual:
 \[\langle p'|j_\mu |p\rangle \langle n'|n\rangle +\langle n'|j_ \mu
|n\rangle\langle p'|p\rangle .\]
The second term in eq.(\ref{decomp}) describes the "real" n-p interaction.
The decomposition in eq.(\ref{decomp}) is relativistic invariant due to the
existence of the one-particle singular invariants of the type $2p_0\delta
(\vec{p}-\vec{p'})$.

\noindent 3. The calculation of $g^{..}_{..}$ as a function of invariant variables
and  nucleon EMFFs. These calculations were done by the methods of
relativistic kinematics (without any $v/c$--expansion). It's also evident
from the two previous points that the nucleon EMFFs,entering to
$g^{..}_{..}$, describe the nucleons on their mass shell.

\noindent 4.The local analyticity of the FFs
\begin{equation}
G^{int}=\lim_{\epsilon ,\eta \rightarrow 0} G^{int}(s\pm i\epsilon ,
s'\mp i\eta,...)
\label{2}
\end{equation}
in the vicinity of the physical region of the variables $s,s'\geq 4M^2\;$.
As we defined above, $s\:\mbox{and}\: s'$ are the squared invariant masses
of n-p system in initial and final states.

\noindent 5. The representation of the n-p scattering matrix $S(s)$ in the physical
region in terms of the Jost matrix $B(s)$:
\begin{equation}
 S(s)=B(s-i\epsilon )B^{-1}(s+i\epsilon ),\qquad s \geq 4M^2\:,
\label{Jost}
\end{equation}
i.e. the solution of the boundary problem
\begin{equation}
 S(s)B_{+}(s)=B_{-}(s),\qquad \qquad s \geq 4M^2
\label{Bound}
\end{equation}
in the scattering theory. As it well known, in the model one-channel case
the the solution of eqs.(\ref{Jost},\ref{Bound}) has the simple form
\[
B(s)=\left(1-\frac{M^2_d-4M^2}{s-4M^2}\right)\cdot \exp \left\{\frac{1}{\pi }
\int^\infty_{4M^2}\frac{\delta (u)du}{s-u}\right\}
\]
where $\delta $ is the phase-shift for one channel $S$--matrix
($ S=\exp(2i\delta )$).
But in the channel of n-p scattering with the deuteron's quantum numbers
the scattering matrix is $2\times 2$ matrix:
\begin{equation}
  S(s) \equiv S \left[ \delta ,\eta ,\varepsilon \right]= \left(
  \begin{array}{ll}
  \cos{2\varepsilon }\cdot e^{2i \delta }& i\sin {2\varepsilon } \cdot
  e^{i( \delta + \eta )}\\
  i\sin {2\varepsilon } \cdot e^ {i( \delta + \eta )} &
  \cos{2\varepsilon }\cdot e^ {2i \eta }
  \end{array}
\right)\;.
\label{four}
 \end{equation}
In this case the solution of the boundary problem (eq.(\ref{Bound})) is
much more complicated, but it appears to be possible to find B in the form
of series with fast convergence (see below). It is essential that every
element of the inverse matrix $B^{-1}$ has a simple pole at the point of
bound state $ s= M^2_d$.

\noindent 6. Solution of the linear conjugation problem for $G^{..}_{..}$. Relative
to variable $s$, for example, the initial equation is
\begin{equation}
\left [ g(s,..)+G(s+i\epsilon ,..) \right ] =
\left [ g(s,..)+G(s-i\epsilon ,..) \right ] \cdot S(s).
\label{initG}
\end{equation}
Following by this way and combining  the solution of eqs.(\ref{Jost},
\ref{initG}) for $s$-variable and the analogous one for $s'$- variable, we
can obtain the integral representation for the relativistic EMFFs of n-p
system in terms  of $g$ and $B$. The resulting formulae are cumbersome
enough and we omit them here.

\noindent 7. Taking the residues at the deuteron poles $s=s'= M^2_d$ we
obtain the final formulae for the DEMFFs (see,for example,
eqs.(\ref{one},\ref{two}).

After this sketch of the method we return to the analysis of RDCFF.
 Among the three DEMFFs the $DCFF$ is the most interesting for us when
investigating the deuteron structure. The point is that the  most
informative features of the deuteron structure --- zero (or "diffractional
 minimum") and prominent secondary maximum in the theoretical  predictions
for DEMFFs --- appear already in the DCFF at not so large values of $Q^2$
($ \sim 20-30 \;fm^{-2}$).  The measurements of $ |G_{cd}(Q^2)|$ for such
$Q^2$ are real [1,2]. On the contrary, the theoretical picture of DM(Q)FFs
is so that indicated above the fine structure in this FFs appears for
relatively high $Q^2$, which hardly can be reached in current and planned
polarization experiments.

        Following \cite{roman} the formulae for RDCFF $G_{cd}(Q^2)$
appear  as
 \begin{eqnarray}
G_{cd}(Q^2)& = & (\rho \tilde{B}^{20} +\tilde{B}^{22})^2 G_{cd}^{00}(Q^2) -
\nonumber\\
&&-(\rho \tilde{B}^{20} + \tilde{B}^{22})(\rho \tilde{B}^{00} +
\tilde{B}^{02}) [G_{cd}^{02}(Q^2) + G_{cd}^{20}(Q^2) ] + \nonumber\\
&&+ (\rho \tilde{B}^{00} + \tilde{B}^{02})^2 G_{cd}^{22}(Q^2)\;,
\label{one}
\end{eqnarray}
\begin{eqnarray}
 G_{cd}^{ll'}&=&\Gamma ^2 \int_{4M^2}^{\infty} \frac{ds\Delta B^\dagger(s)}
 {s - M^2_d}\int_{s_1(s,t)}^{s_2(s,t)} \frac{ds' g_c(s,s't) \Delta B(s')}
{s'-M^2_d}, \nonumber\\
s_{2,1}&=&2M^2 + \frac{1}{2M^2}(2M^2 -t)\cdot(s-2M^2) \pm \nonumber\\
&&\pm \frac{1}{2M^2} \sqrt{(-t)(4M^2-t)s(s-4M^2)}\;.
\label{two}
\end{eqnarray}

       In  eq.(~\ref{one})  $\rho  $    is  the  constant,  which  describes
 the mixing of two $n-p$ states  with different orbital moments ($l=0$   and
 $l=2$) in the point  of  the bound  state, i.e., the  deuteron. This constant
 is  defined   by    the    correspondence    principle.    Analyzing    the
nonrelativistic limits of  eqs.(\ref{one},\ref{two}), we can  prove that
$\rho $ appears to be the standard  asymptotic $D/S$ --ratio  of the
radial deuteron   wave functions,   so $   \rho =   0.0277$ (numerical
calculations show  that the dependence of DCFF on the variation  of $\rho $
 is very  weak). All   four elements  of the  matrix $ \tilde{B}^{ll'}(s)$
($ l,l' = 0,2$) are taken  at  the  bound  state  point  $  s=M_d^2$  ($M_d
=  2M -\varepsilon  $, where  $M_d,M$    being  deuteron and nucleon
masses, and $\varepsilon $--  being the deuteron  binding energy).

        In eq.(\ref{two}) $\Gamma ^2$ is the normalization constant,
 which is calculated from the condition  $G_{cd}(0) =1$. Matrix
 functions $\Delta B^{ll'}(s) = B^{ll'}(s + i \varepsilon ) - B^{ll'}
 (s-i\varepsilon )$ are the discontinuities of the Jost matrix $B(s)$.
The reduced Jost matrix $\tilde{B}$ in eq.(~\ref{one}) is the solution of
the same as for $B$, eq.(~\ref{Bound}) with the scattering matrix
$ \tilde{S}\equiv S \left [\tilde{\delta }, \tilde{\varepsilon},
\tilde{ \eta }\right ]$. The reduced phase  shifts $\tilde{\delta },
\tilde{\varepsilon}, \tilde{ \eta } $ have the auxiliary nature and
describe the n-p scattering without bound state ($ \tilde{ \delta }
(E\rightarrow 0) \rightarrow 0$). Roughly speaking, matrix $\tilde{B}$
is the rest of matrix $B$ after taking the residue at the point of the
deuteron's pole. The formulae for $\tilde{B}$ and $B$ in
terms of $n-p$ phase shifts are cumbersome and are summarized in Appendix
B. All  relativistic aspects of the two--nucleon problem are contained
in $G^{ll'}$-- matrix.

As was mentioned above, the matrix functions $ g^{ll'}_c(s,s',t)$ of three
variables are the relativistic CFF of the unconnected part of the matrix
element of EM current $\langle n'p'|j_\mu |np\rangle$. The results of the
calculations of $ g^{ll'}_c$ are given in Appendix C. It is interesting to
 note that in general case in the relativistic region $ g^{ll'}_c$-- functions
are not factorized in $ s,s'$ variables, whereas in the nonrelativistic
limit such factorization takes place. It means that in the framework of
  the used relativistic approach \cite{tro}--\cite{roman} it is impossible to
introduce some kind of concept of relativistic deuteron wave function.

  The $n-p$ phase shifts were taken from the recent analysis of Virginia
Tech group~\cite{Virginia} and is shown in Fig.~\ref{fig2}. This analysis
was made in the energy range \mbox{$ 0 < E_{lab} \leq 1300 $ MeV}.
  Extrapolation to higher energies is not so  important for the
  calculations of $G_{Cd}$  for the small and intermediate values of $Q^2$
  (see below).  The only essential circumstance is  that $^3S_1$--phase
  shifts change  sign from positive to negative and have the minimum near
  the energy $E_{lab} \sim 1 $GeV, then go to zero in accordance with the
  Levinson's theorem. Any realistic $n-p  $ $ ^3S_1$ -- phase shift
  analysis has such a behavior.  Two other states ($^3D_1$ and $^3
  \varepsilon _1$ ) give a relatively small contribution to $G_{Cd}$.

      For the calculations of $G_{Cd}$ we used (as a first step) the
  simplest choice of the nucleon form factors: $ G_{Ep} = (1+Q^2/18.23\;
  fm^{-2})^{-2}, G_{Mp}/\mu _p = G_{Mn}/\mu _n = G_{Ep},\; G_{En} \equiv 0
  $ for all $ Q^2$.

      The result of the calculations are presented in Fig.~\ref{fig3}. Our
  brief conclusions are as follows.

  \noindent 1. The appearence  of zero and secondary maximum in
  $|G_{Cd}(Q^2)|$ at intermediate values of $Q^2$ is the direct consequence
   of the change of sign of the experimental $^3S_1$-- phase shifts at
  intermediate energies.  It is easy to calculate that the model's  $\delta
  (E)$, which decreases monotonically with $E$ and is always positive
  ($\delta (E) > 0 $ for all $E$), immediately leads to monotonically
  decreasing with $Q^2$ values of $|G_{Cd}(Q^2)|$  without any fine
  structure.

 \noindent 2.  Almost up to  the point of zero ($Q^2 \simeq 20\; fm^{-2})$
  of $|G_{Cd}(Q^2)|$  the total relativistic correction (TRC), {\it i.e.},
  the difference between $G_{Cd}$ calculated relativistically (1,2) and its
  nonrelativistic limit, is positive and appears to be not small.  For
  example, for $Q^2 \simeq 14\; fm^{-2} $ it reaches the value of 25\%.

\noindent  3. TRC becomes large in the region of the secondary maximum
of $|G_{Cd}(Q^2)|$, increasing the magnitude of the form factor.

\noindent 4. The obtained results are consistent with the available data on
  $G_{Cd}$ from MIT-Bates \cite{bates}. New data from Jefferson Lab
E--94-018 \cite{betsy} are extremely important to test the proposed
relativistic approach in the region of higher transferred momenta, where
relativistic corrections appear to be significant.

    We would like to make the following comments to the obtained results.
 Firstly, as it was mentioned above, for the fixed till
        $ E_{lab} = 1300 \mbox{ MeV} \;^3S_1$--phase shifts set of ref.
  \cite{Virginia} the character of extrapolation to high energies is not
  essential when calculating the RDCFF. Two other, than in Fig.\ref{fig2}
  possible extrapolations of $\delta (E\rightarrow \infty)\;$ are shown
  in Fig.\ref{fig4}. Their foundations are the following. If we include
the so-called forbidden bound states in NN system (possible in some kinds
  of theories) in the formulation of Levinson's theorem, then the
  immediate consequence for $\delta (E)\;$ would be $\delta (E\rightarrow
  \infty)\rightarrow -\pi\;$ (see,for example, ref.\cite{VGN}.
The second extrapolation is connected with the phase shift analysis of
proton--proton scattering for the several GeV region (ref.\cite{Virginia}).
It is natural to think that for high energies the contribution of Coulomb
effects are small, so n-p and p-p phase shifts in the equal spin states
are very close to each other. The results of the calculations of
$|G_{Cd}(Q^2)|$ with the phase shifts from Fig.\ref{fig4} are shown in
Fig.\ref{fig5}. It is evident that the difference between these three
curves is negligible. But when we turn to other phase shifts, which are
differing from the results of ref.\cite{Virginia} in the intermediate
energy region, $ E_{lab} < 1 \mbox{ GeV} $, the situation change
essentially. The dependence of $|G_{Cd}(Q^2)|$ structure on the choice of
  different sets of experimental $n-p$ phase shifts  available from the
literature (ref.\cite{band}) is strong enough. Possible variation of
$\delta , \varepsilon , \eta $ may shift the position of zero in
   $|G_{Cd}(Q^2)|$ from the indicated point $Q^2 = 20\;fm^{-2}$ to the
  point $Q^2 = 16\;fm^{-2}$ or to the point $Q^2 = 23\;fm^{-2}$. At the
 same time the secondary maximum is located in the interval $26\leq Q^2
  \leq 32 \; fm^{-2}$, and its height may change by a factor of seven. We
  can see that for improving our understanding of $|G_{Cd}(Q^2)|$  it would
  be  desirable to have a more detailed phase shifts analysis  of $n-p$
  scattering in triplet channel in intermediate energy region $E_{lab} \leq
  1 $ GeV.  Secondly, let us indicate the dependence of $|G_{Cd}(Q^2)|$ on
  the possible choice of nucleon EM form factors. Since the uncertainties
of $G_{Ep}(Q^2)$ in the considered range of $Q^2$ are very small, the main
effect in $|G_{Cd}(Q^2)|$ may be caused only by variation of $G_{En}(Q^2)$.
It seems to be generally accepted that the maximal deviation of
$G_{En}(Q^2)$  from the zero-value  approximation $G_{En} \equiv 0$ is
given by known formula $G_{En}(Q^2) = - \mu _n\tau G_{Ep}(Q^2)$, where $\mu
_n = -1.91$ is the neutron anomalous  magnetic moment and $\tau =
Q^2/4M^2$. The results of the calculations of $|G_{Cd}(Q^2)|$ with this
nonzero values of $G_{En}(Q^2)$  are shown in fig.2.  One can see that the
effect is sizable and the contributions of relativistic effects and nonzero
$G_{En}$ have a similar behavior.

   Finally, we show for comparison in Fig.\ref{fig3} the results of
calculation of $G_{Cd}$ in a relativistic approach developed in
ref.~\cite{wally}. It may be seen that zero of $|G_{Cd}(Q^2)|$ predicted in
ref. ~\cite{wally} is shifted to the lower values of $Q^2$ and the height
of the secondary maximum is approximately the same as in our calculations.

      Here we restricted ourselves only to the discussion of the deuteron
charge form factor $G_{Cd}$.  Even in this case we omitted such
interesting questions as an  analytical representation of relativistic
corrections in different orders in $(v/c)^2$, the new representation for
realistic deuteron wave functions, the role of relativistic rotation of
nucleon spins and orbital momentum $l=2$ in the deuteron, the problem of
extraction, using the present approach, of $G_{En}(Q^2)$ for ultralow
values of $Q^2$ from experimental data on elastic ed--scattering, and
contributions from meson--exchange currents.  It would also be interesting
  to perform a detailed comparison of the present approach with other
relativistic approaches to the description of deuteron structure.

All these questions, as well as the calculations of the deuteron magnetic
and quadrupole form factors will be discussed in forthcoming publications.
\bigskip

{\bf Acknowledgements}.

We would like to thank F. Gross, J.W. Van Orden, and D. Sprung for useful discussions. The work of A.A. was supported by
  the US Department of Energy under contract DE--AC05--84ER40150.
\end{sloppypar}

\appendix
 \section{Kinematic variables.}

      By definition $s$ is the invariant mass of $n-p$ system squared:
  \[s= (p_n + p_p)^2_\mu \:. \]
  In laboratory (LS) and center-of-mass (CMS)  systems we have
   \[ s = 4M^2+2E = 4M^2 +4p^2 \:,\]
where $E$ is the nucleon energy in LS and $p$ is the magnitude of the nucleon 3--momentum in CMS.

      $Q^2$ is the magnitude of the 4--momentum transfer squared:
\[ Q^2 \equiv -q^2_\mu \equiv -t > 0 \:.\]

\section{Jost matrices $B,\tilde{B}$.}

      The formulae for pairs ($S,B$) and ($\tilde{S},\tilde{B}$) have the
most convenient form in the $p$--plane:
\[S(p)B_{+}(p) = B_{-}(p)\:,\mbox{   } -\infty < p < \infty \:,\]
where $ S\equiv S[\delta (p), \eta (p), \varepsilon (p)]$, see
eq.(\ref{four}). Let us  introduce two new matrices $\tilde{S}$ and
$\tilde{B}$:

   \[ \begin{array}{rlr}
        \tilde{B}_{\pm}(p)& = R(\mp p)B_{\pm}(p), &   \\
          R(p) &= I-{2i\alpha \over(p+i\alpha )(1+\rho ^2)}\cdot
         \left( \begin{array}{lr}
                    1&-\rho \\
                   -\rho &\rho ^2
                 \end{array}
         \right), &(\alpha ^2 = M\varepsilon) .
        \end{array}
     \]

   Now  the equation for $\tilde{B}$ has the form
\begin{equation}
  \left\{ \begin{array}{rlr}
       \tilde{S(p)}\tilde{B}_{+}(p)=&\tilde{B}_{-}(p),&-\infty<p<\infty\;,\\
       \tilde{S}(p)=&R(p)S(p)R^{-1}(-p)\equiv
       \tilde{S}[\tilde{\delta }, \tilde{ \eta },\tilde{\varepsilon }].&
       \end{array}
  \right.
  \label{b1}
  \end{equation}

   The last equation defines the reduced phase shifts $\tilde{\delta },
   \tilde{\varepsilon },\tilde{\eta}$ as functions of input experimental
   phase shifts $\delta ,\varepsilon ,\eta $.

   The solution of eq.(\ref{b1}) was found in ref.\cite{Muz} in the form of
   series
   \[
     \tilde{B}_{\pm}(p) = \tilde{B}_{\pm,0}(p)\cdot[I+\sum_{m=1}^{\infty}
     \tilde{B}_{\pm,m}(p)],
    \]
    where
  \[
   \tilde{B}_{\pm,0}(p)=
     \left(
          \begin{array}{lr}
                 \varphi _1(p)e^{\mp \tilde{ \delta }(p)}&0\\
                  0&\varphi _2(p) e^{\mp \tilde{ \eta }(p)}
           \end{array}
      \right)\:,
     \]

    \[
    \varphi _1(p)= \exp[-\frac{1}{\pi }  V.P. \int _{-\infty}^{\infty}
        {\tilde \delta (p')dp' \over p'-p}],
     \]
    \[
    \varphi _2(p)= \exp[-\frac{1}{\pi }  V.P. \int _{-\infty}^{\infty}
        {\tilde \eta (p')dp' \over p'-p}],
     \]

\begin{equation}
  \tilde B_{\pm,m}(p) =\frac{1}{2\pi i} \int_{-\infty}^{\infty} { dp'
   \over p-p' \pm i0} \cdot \sum_{n=1}^m G_n(p')\tilde B_{+,0}(p')
   [\tilde B_{+,m-n}(p')]^{1-\delta _{mn}}.
\label{b2}
\end{equation}

  In eq.(\ref{b2}) for odd $n$
  \[
    G_n(p) =i(-1)^{\frac{n-1}{2}}\cdot \frac{1}{n!}\cdot
                 [2 \tilde \varepsilon(p)]^n\cdot
                \left( \begin{array}{ll}
                       0& e^{i(\tilde \delta +\tilde \eta)} \\
                       e^{i(\tilde \delta +\tilde \eta)} &0
                        \end{array}
                 \right )
   \]
 and  for even $n$
  \[
    G_n(p) =i(-1)^{\frac{n}{2}}\cdot \frac{1}{n!}[2 \tilde \varepsilon
   (p)]^n\cdot \left( \begin{array}{ll}
                       e^{2i\tilde \delta } &0\\
                       0& e^{2i\tilde \eta}
                        \end{array}
                 \right ),
   \]
  $\delta _{mn}$ is the Kroneker delta.

\section{$g^{ll'}_c$--matrix.}
      In terms of invariant variables $s,s',t$ and the nucleon EM form factors the
matrix elements have the form:
\begin{eqnarray*}
\lefteqn{g_c^{00}(s,s',t) = }\\
&&g(s,s',t)[g_1(s,s',t)(\cos\alpha _1 \cos\alpha _2 - \frac{1}{3}
  \sin\alpha_1\sin\alpha _2)\cdot G_{EN}^s(Q^2) +\\
&&+\frac{1}{2M} g_2(s,s',t)\cdot (\frac{1}{3}\sin \alpha _1 \cos\alpha _2 -
\cos \alpha _1 \sin \alpha _2)\cdot G^s_{MN}(Q^2)] \;,
\end{eqnarray*}
\begin{eqnarray*}
\lefteqn{g_c^{02}(s,s',t) = }\\
&&g(s,s',t)\{ g_1(s,s',t)(-\sqrt{2} P_{20}\cos\alpha _1\sin\alpha _2 +
\frac{1}{\sqrt 2}P_{21} \sin\alpha _1 \cos\alpha _2 ) \cdot G^s_{EN} \\
&&-\frac{1}{2M} g_2(s,s',t) (\sqrt 2 P_{20} \sin\alpha _1 \cos\alpha _2 +
\frac{1}{\sqrt 2}P_{21} \cos \alpha _1 \sin \alpha _2 ) G^s_{MN}  \}, \\
\end{eqnarray*}
\[  g_c^{20}(s,s',t) =  g_c^{02}(s',s,t),  \]

\begin{eqnarray*}
\lefteqn{g_c^{22}(s,s',t) = }\\
&&g(s,s',t) \Bigl\{ g_1(s,s',t)
  \Bigl[(\frac{1}{3} P_{21} P'_{21} + \frac{2}{3}
   P_{20}P'_{20})\cos(\alpha _1 - \alpha _2) + \\
&& + ( \frac{1}{12} P_{22}P'_{22} + \frac{1}{3} P_{20}P'_{20})\cos\alpha _1
\cos\alpha _2 +  \\
&&+\Bigl(
          \frac{1}{12}(P_{22}P'_{21} - P_{21}P'_{22}) +
          \frac{1}{2}(P_{21}P'_{20} -P_{20}P'_{21})
    \Bigr)\sin( \alpha _1 - \alpha _2 ) - \\
&& - \frac{1}{6}(P_{22}P'_{20} + P_{20}P'_{22}) \sin\alpha _1 \sin\alpha _2
  \Bigr ]\cdot G^s_{EN} - \frac{1}{2M}g_2(s,s',t)\cdot  \\
&& \Bigl[
 \frac{1}{12}\bigl((P_{21}P'_{22} - P_{22}P'_{21}) +
 \frac{1}{2} (P_{20}P'_{21}-P_{21}P'_{20})\bigr) \cdot \\
&&\cos(\alpha _1-\alpha  _2)-( \frac{1}{12} P_{22}P'_{22} + \frac{1}{3}
P_{20}P'_{20})\cos\alpha _1\sin\alpha _2 -\\
&&-\frac{1}{6}( P_{22}P'_{20} -
 P_{20}P'_{22}) \sin\alpha _1 \cos\alpha _2 + \\
 && + ( \frac{1}{3} P_{21}P'_{21}+\frac{2}{3}P_{20}P'_{20})
  \sin ( \alpha _1 - \alpha _2)
  \Bigr] \cdot G^s_{MN}
  \Bigr\}\;,
 \end{eqnarray*}

where
\[
\begin{array}{rl}
  g(s,s',t) = & {g_1(s,s',t)\cdot(-t) \over\sqrt{(s-4M^2)(s'-4M^2)}} \cdot
   \frac{1}{[\lambda (s,s',t)]^{3/2}} \cdot \frac{1}{\sqrt{1+\tau }} ,\\
  g_1(s,s',t) = & s+s'- t,\\
  g_2(s,s',t) = & \Bigl[(-1)\bigl(M^2\lambda (s,s',t)+s s' t\bigl)
                   \Bigr]^{1/2},\\
\lambda (s,s',t) = & s^2+s'^2+t^2 -2(ss'+st+s't)  .
\end{array}
\]

      $P_{lm}$ are the Legendre polynomials, $P_{lm}\equiv P_{lm}(x)$ and
$P'_{lm}\equiv P_{lm}(x')$, where

\[
\begin{array}{rl}
x(s,s',t)=& \frac{\sqrt{s'(s'-s-t)}}{\sqrt{(s'-4M^2)\lambda (s,s',t)}},\\
 x'(s,s',t)=& -x(s',s,t).
\end{array}
\]
The angles $\alpha _1,\alpha _2$ of the relativistic rotation of nucleon
spins in deuteron are
\[
\begin{array}{ll}
\alpha _1 =& \arctan{g_2(s,s',t) \over M\bigl[(\sqrt{s} +\sqrt{s'})^2 -
              t \bigr] +\sqrt{ss'}(\sqrt{s} +\sqrt{s'} +2M)},\\
\alpha _2=& \arctan{g_2(s,s',t) (\sqrt{s} +\sqrt{s'} +2M) \over
           M(s+s'-t)(\sqrt{s} +\sqrt{s'}+2M) +\sqrt{ss'}( 4M^2-t)}.
\end{array}
 \]
     $\tau =Q^2/4M^2$ ; $G^s_{E,MN} =\frac{1}{2}(G_{E,Mp} + G_{E,Mn})$
are the nucleon isoscalar charge and magnetic form factors.

\newpage
\begin{figure}[t]
\let\picnaturalsize=N
\def\picsize{3in}
\def\picfilenamea{fig1.eps}
\ifx\nopictures Y\else{\ifx\epsfloaded Y\else\input epsf \fi
\let\epsfloaded=Y
\centerline{
\ifx\picnaturalsize N\epsfxsize \picsize\fi \epsfbox{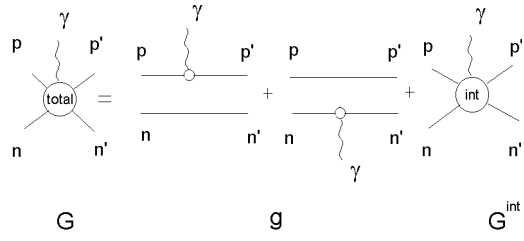}}}
\fi
\caption{Separation of the unconnected part of the two-nucleon electromagnetic current.}
\label{fig1}
\end{figure}
\newpage

\begin{figure}[t]
\let\picnaturalsize=N
\def\picsize{4in}
\def\picfilenamea{fig2.eps}
\ifx\nopictures Y\else{\ifx\epsfloaded Y\else\input epsf \fi
\let\epsfloaded=Y
\centerline{
\ifx\picnaturalsize N\epsfxsize \picsize\fi \epsfbox{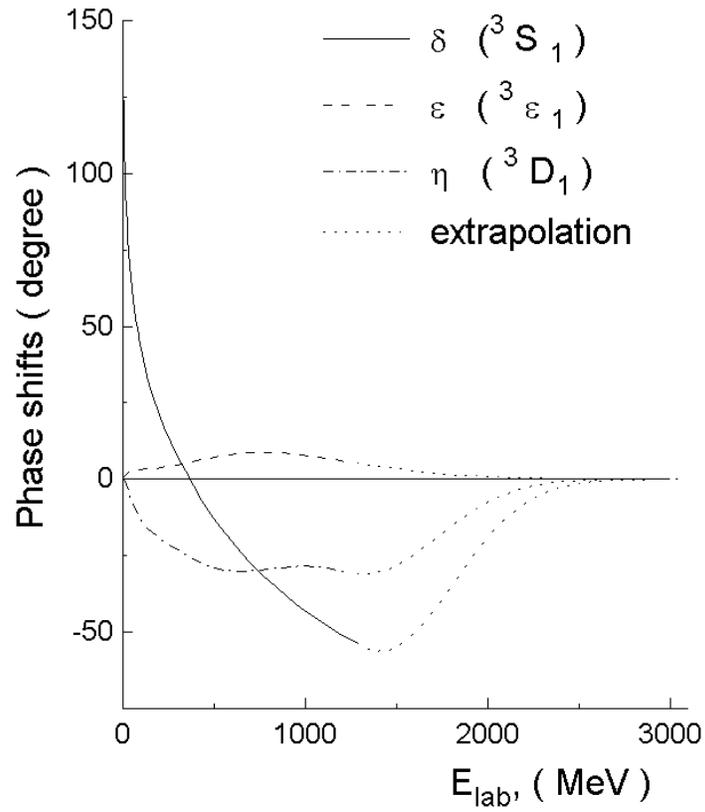}}}
\fi
\caption{Neutron--proton phase shifts from ref.\protect \cite{Virginia} and
their extrapolation.}
\label{fig2}
\end{figure}

\begin{figure}[t]
\let\picnaturalsize=N
\def\picsize{4in}
\def\picfilenamea{fig3.eps}
\ifx\nopictures Y\else{\ifx\epsfloaded Y\else\input epsf \fi
\let\epsfloaded=Y
\centerline{
\ifx\picnaturalsize N\epsfxsize \picsize\fi \epsfbox{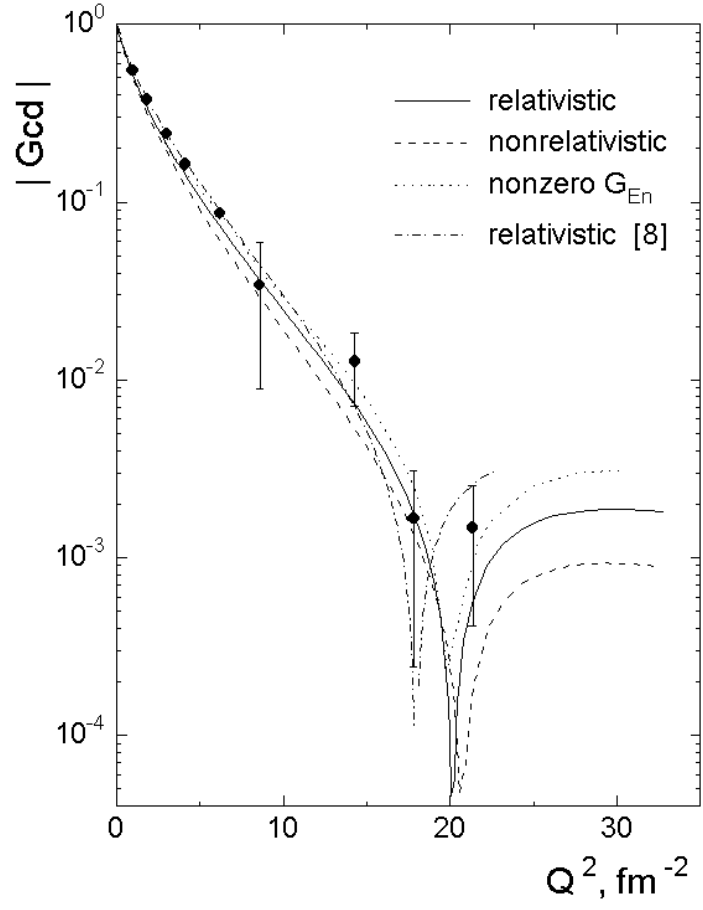}}}
\fi
\caption{Relativistic deuteron charge form factor and its nonrelativistic
limit. RDCFF with nonzero values of $G_{En} = -\mu _n\tau G_{Ep}$ is
also shown. For comparison DCFF in relativistic approach of
\protect \cite{wally}  and experimental results from
MIT--Bates \protect \cite{bates} are given, too.}
\label{fig3}
\end{figure}
\begin{figure}[t]
\let\picnaturalsize=N
\def\picsize{4in}
\def\picfilenamea{fig4.eps}
\ifx\nopictures Y\else{\ifx\epsfloaded Y\else\input epsf \fi
\let\epsfloaded=Y
\centerline{
\ifx\picnaturalsize N\epsfxsize \picsize\fi \epsfbox{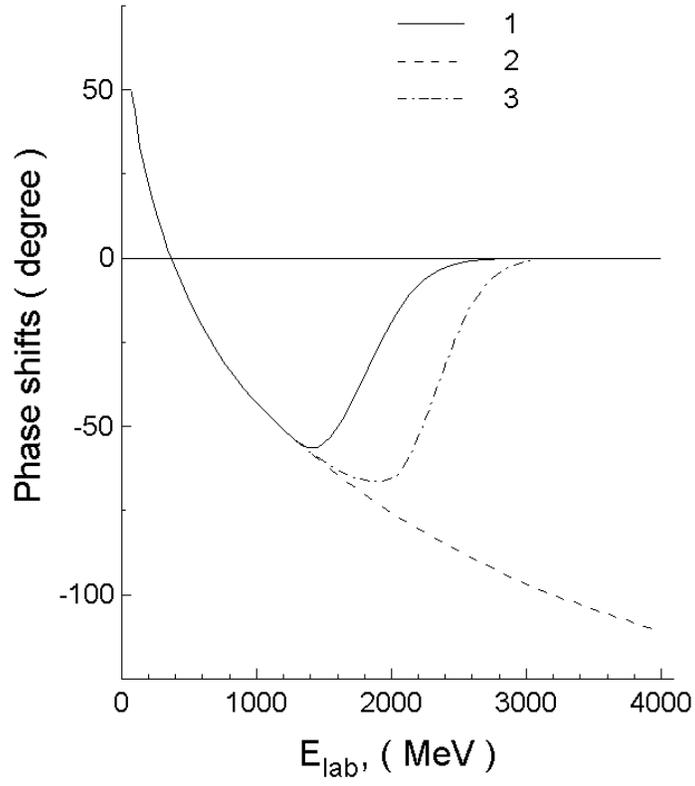}}}
\fi
\caption{The various extrapolations of $^3S_1$--phase shifts to
high energies. Curve 1 is the same as in Fig.2. The curves
2,3 are described in the text.}
\label{fig4}
\end{figure}
\begin{figure}[t]
\let\picnaturalsize=N
\def\picsize{4in}
\def\picfilenamea{fig5.eps}
\ifx\nopictures Y\else{\ifx\epsfloaded Y\else\input epsf \fi
\let\epsfloaded=Y
\centerline{
\ifx\picnaturalsize N\epsfxsize \picsize\fi \epsfbox{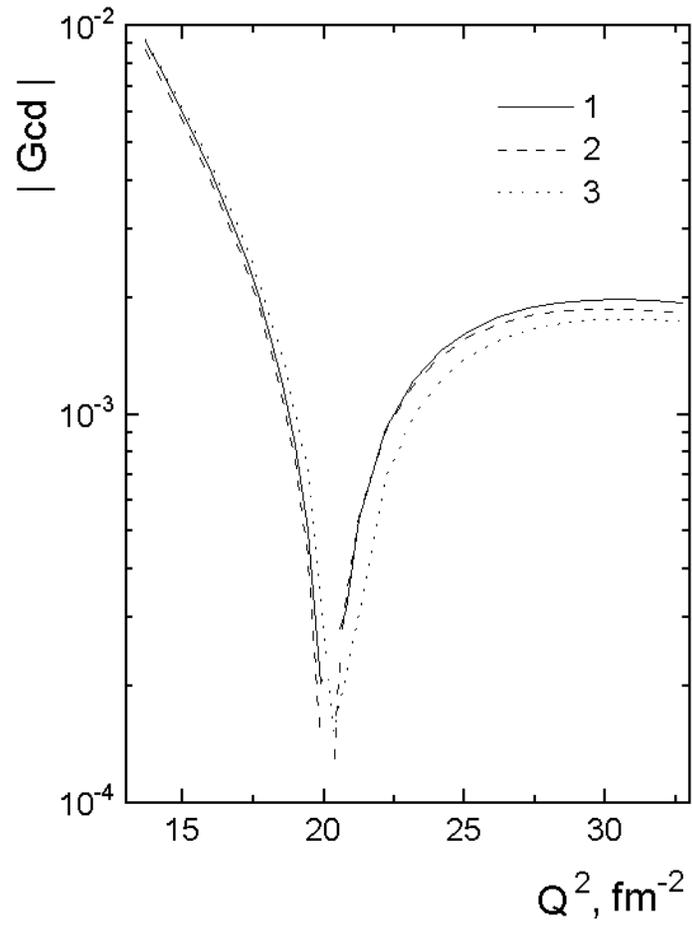}}}
\fi
\caption{ RDCFF for the phase shifts from Fig.4.}
\label{fig5}
\end{figure}

\end{document}